\documentclass[a4paper]{IEEEtran}

\usepackage{amsmath,amssymb}

\usepackage{tikz}
\usetikzlibrary{shapes}
\usetikzlibrary{circuits.ee.IEC}
\usetikzlibrary{colorbrewer}
\usepackage{pgfplots}
\usepgfplotslibrary{colorbrewer}
\usepgfplotslibrary{groupplots}

\pgfplotsset{compat=1.14}

\usepackage{cite}

\begin{document}

\title{Experimental Demonstration of Learned Time-Domain Digital Back-Propagation}

\author{Eric~Sillekens, Wenting~Yi, Daniel~Semrau, Alessandro~Ottino, Boris~Karanov,
Sujie~Zhou, Kevin~Law, Jack~Chen, Domaniç~Lavery, Lidia~Galdino, Polina~Bayvel, and~Robert~I.~Killey%
\thanks{Eric~Sillekens, Wenting~Yi, Daniel~Semrau, Alessandro~Ottino, Boris~Karanov, Domaniç~Lavery, Lidia~Galdino, Polina~Bayvel, and~Robert~I.~Killey are with the Optical Networks Group, Dept. Electronic \& Electrical Engineering, UCL, London WC1E 7JE, U.K.}%
\thanks{Sujie~Zhou and Jack~Chen are with Huawei Chengdu Research Institute U1, Chengdu, Sichuan Province, P.R. China, Postal Code 611731}%
\thanks{Kevin~Law is with Huawei Base G6, Dongguan, Guangdong Province, P.R. China, Postal Code 523808}%
}

\maketitle

\begin{abstract}
We present the first experimental demonstration of learned time-domain digital back-propagation (DBP), in 64-GBd dual-polarization 64-QAM signal transmission over 1014 km. Performance gains were comparable to those obtained with conventional, higher complexity, frequency-domain DBP.%
\end{abstract}

\section{Introduction} 
The non-linear fibre channel has a limited capacity due to increasing nonlinear signal distortions with increasing transmission power, leading to a peak in achievable information rate (AIR) for a fixed bandwidth. One approach to increase this maximum AIR is to mitigate the non-linear distortion. This can be achieved with digital signal processing (DSP) by solving the differential equation that describes the non-linear fibre response backwards, using the received signal as the initial condition. This method, known as split-step Fourier method (SSFM) based digital back-propagation (DBP)\cite{Ip2008Compensation}, has been shown to allow increased data throughput and transmission reach \cite{Galdino2017OnTheLimits}.

A significant drawback of the DBP technique is its computational complexity, making it challenging to implement in real-time systems. In conventional frequency-domain DBP (FD-DBP), dispersion compensation is performed in the frequency domain and nonlinear phase shifts are corrected in the time domain, requiring repeated conversions of the signal between the time and frequency domains using fast Fourier transforms. This leads to high computational complexity, particularly when small step-sizes, and hence a large number of steps, are used to achieve high accuracy. To reduce complexity, both dispersion and nonlinearity could be compensated in the time-domain (TD-DBP), with the dispersion compensation being carried out with tap-and-delay finite impulse response (FIR) filters \cite{Fougstedt2017Time-Domain}. However, low-order FIR filters are fundamentally unable to accurately compensate small amounts of dispersion. An approach to overcome this drawback was proposed in \cite{Haeger2018Nonlinear,Haeger2018Deep}, and involves
applying machine-learning techniques to optimize the combined response of all the cascaded filters. The approach leverages the similarities between time-domain DBP and deep feed-forward neural networks; in both structures, linear filters and nonlinear functions are interleaved. The recent rapid advances in algorithms, and readily available software packages, allow implementation of these algorithms for the optical transmission channel.

In this work, we experimentally demonstrate, for the first time, learned time-domain digital back-propagation. First, the method of training the required time-domain filter weights is explained. Next, the performance of the learned TD-DBP is assessed for 4-channel 64~GBd polarisation division multiplexing (PDM)-64QAM transmission over 1014 km, and compared with the performance of conventional frequency-domain DBP. Finally, the resulting filter tap weights and frequency response of the FIR filters are analysed. We observed performance improvements over linear compensation comparable to those obtained using the conventional FD-DBP implementation.

\section{Digital Back-Propagation} 

DBP implements the non-linear Schrödinger equation (NLSE) for each step in two parts. For a step starting at distance $z$ along the fibre, firstly, the chromatic dispersion and loss are applied, in the frequency domain (in the case of the conventional FD-DBP implementation) as a linear operator, and the non-linear phase shift is performed in the time domain, using the respective transformations;%
\begin{align}%
    E(\omega, z+\Delta z) &= e^{\alpha \Delta z}e^{jK(\omega T)^2}E(\omega, z) \\
    E(t,z+\Delta z) &= e^{-j\gamma \Delta z|E_z(t)|^2}E(t,z),
\end{align}%
where $\alpha$ is fibre loss, $\Delta z$ is fibre step length, $K=\frac{\beta_2\Delta z}{2T^2}$, $\omega$ angular frequency, $T$  sampling period, and $\beta_2$ group velocity dispersion, $\gamma$ the nonlinearity coefficient and $|E_z(t)|^2$ the normalised, step-averaged, instantaneous optical power. 

In the time-domain DBP approach, the chromatic dispersion part of each step is applied using a time-domain finite impulse response (FIR) filter (tap-and-delay filter). The full-band least-squares FIR filter design from \cite{Eghbali2014Optimal} could be used with a total number of taps given by $N_c \le 2 \lfloor 2 \pi K \rfloor +1$. However, as described in \cite{Haeger2018Nonlinear}, if the filter tap weights given by \cite[Eq.~(13)]{Eghbali2014Optimal} are used for the multiple cascaded low-order FIR filters in the DBP, the ripples introduced into the frequency response result in large performance penalties, negating the gains achieved through the non-linearity mitigation. The solution proposed in \cite{Haeger2018Nonlinear} is to update all the FIR filter weights simultaneously using algorithms which have been developed to update the weights in deep feed-forward neural networks. The method consists of implementing the dispersion as a convolutional layer and the fibre phase shift as a non-linear activation function. 

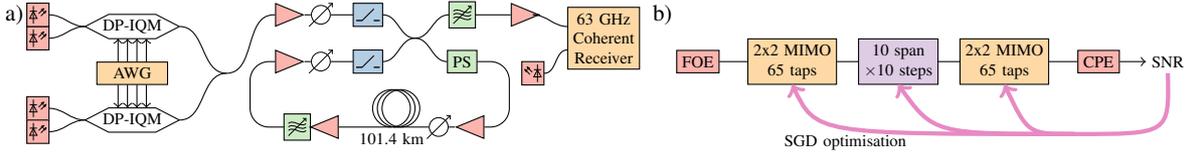
\begin{figure*}[bt]
    \centering
    \newcommand{\lazer}{\begin{tikzpicture}[circuit ee IEC]
    \draw (0,-0.75) to [diode={light emitting}] (0,0);
    \end{tikzpicture}}
\newcommand{\drawvoa}{\begin{tikzpicture}
    \draw (0,0)  circle(0.20cm);
    \draw[->] (-0.25cm,-0.25cm) -- (0.25cm,0.25cm);
    \end{tikzpicture}}
\newcommand{\drawbpf}{\begin{tikzpicture}
    \draw (-0.25cm,0) to [out=45,in=225] (0.25cm,0);
    \draw (-0.25cm,-0.1cm) to [out=45,in=225] (0.25cm,-0.1cm);
    \draw (-0.25cm,0.1cm) to [out=45,in=225] (0.25cm,0.1cm);
    \draw[->] (-0.25cm,-0.25cm) -- (0.25cm,0.25cm);
    \end{tikzpicture}}
\newcommand{\drawaom}{\begin{tikzpicture}[circuit ee IEC]
    \draw (0,0) to [make contact={}] (1,0);
    \end{tikzpicture}}

\begin{tikzpicture}[scale=0.62,transform shape,
circuit ee IEC,
circuit declare symbol=amp,
set amp graphic={draw,shape=isosceles triangle,minimum size=3mm,fill=Pastel1-A}]

\node[font={\Large}] at (-2.5,1) {a)};

\node[draw,xscale=-0.5,yscale=0.5,fill=Pastel1-A] (l1) at (-2,0.5) {\lazer };
\node[draw,xscale=-0.5,yscale=0.5,fill=Pastel1-A] (l2) at (-2,1) {\lazer };

\node[draw,xscale=-0.5,yscale=0.5,fill=Pastel1-A] (l3) at (-2,-1) {\lazer };
\node[draw,xscale=-0.5,yscale=0.5,fill=Pastel1-A] (l4) at (-2,-1.5) {\lazer };
\node[draw,fill=Pastel1-E,minimum width=1.5cm] (awg) at (-0,-0.25) {AWG};

\node[shape=regular polygon,
      regular polygon sides=6,draw,yscale=0.3,minimum size=2cm] (mzm1) at (-0,0.75) {};
\node at (mzm1) {DP-IQM};
\node[shape=regular polygon,
      regular polygon sides=6,draw,yscale=0.3,minimum size=2cm] (mzm2) at (-0,-1.25) {};
\node at (mzm2) {DP-IQM};

\draw (l1.west) to[in=180,out=0] (mzm1.west);
\draw (l2.west) to[in=180,out=0] (mzm1.west);
\draw (l3.west) to[in=180,out=0] (mzm2.west);
\draw (l4.west) to[in=180,out=0] (mzm2.west);

\draw[->] ([xshift=1mm]awg.north) -- ([xshift=1mm]mzm1.south);
\draw[->] ([xshift=-1mm]awg.north) -- ([xshift=-1mm]mzm1.south);
\draw[->] ([xshift=3mm]awg.north) -- ([xshift=3mm]mzm1.south);
\draw[->] ([xshift=-3mm]awg.north) -- ([xshift=-3mm]mzm1.south);

\draw[->] ([xshift=1mm]awg.south) -- ([xshift=1mm]mzm2.north);
\draw[->] ([xshift=-1mm]awg.south) -- ([xshift=-1mm]mzm2.north);
\draw[->] ([xshift=3mm]awg.south) -- ([xshift=3mm]mzm2.north);
\draw[->] ([xshift=-3mm]awg.south) -- ([xshift=-3mm]mzm2.north);

\coordinate (polmux) at (2,-0.25);

\draw (mzm1.east) to[in=180,out=0] (polmux.west);
\draw (mzm2.east) to[in=180,out=0] (polmux.west);

\coordinate (amptx) at(3,1);
\node[inner sep=-0.05cm] (voatx) at (4,1) {\drawvoa};
\draw (polmux.east) to[in=180,out=0] (amptx)  to[amp={near start}] (voatx);

\node[draw,xscale=0.5,fill=Pastel1-B] (aomtx) at (5,1) {\drawaom};
\draw (voatx) -- (aomtx);

\coordinate (loopx) at (6,0.5);

\node[draw,fill=Pastel1-C] (ps) at (7,0) {PS};


\node[inner sep=-0.05cm] (voaltx) at (6.5,-1.4) {\drawvoa};

\draw[rounded corners=0.25cm] (ps) --(8,0) -- (8,-1.4)--(7.5,-1.4);

\draw (7.5,-1.4) to[amp={near start}] (voaltx);

\coordinate (loopspool) at (5.5,-1.4);

\draw (voaltx) -- ([xshift=1]loopspool);
\draw (5.5,-1) circle(0.4);
\draw (5.6,-1) circle(0.4);
\draw (5.7,-1) circle(0.4);

\node[draw,scale=0.75,fill=Pastel1-C] (wss) at (3.5,-1.4) {\drawbpf};
\draw (loopspool) to[amp={near end}]  (wss);

\coordinate (toploop) at (5.5,0);
\node[inner sep=-0.05cm] (voalrx) at (4,0) {\drawvoa};
\node[draw,xscale=0.5,,fill=Pastel1-B] (aomlrx) at (5,0) {\drawaom};
\draw[rounded corners=0.25cm] (wss) -- (2.5,-1.4) -- (2.5,0) --(3,0);
\draw (3,0) to[amp={near start}] (voalrx) -- (aomlrx);

\node[draw,scale=0.75,fill=Pastel1-C] (bpf) at (7,1) {\drawbpf};

\draw (bpf) to[amp={near end}] (8.5,1)  coordinate (rxamp);

\draw (aomtx) to[in=180,out=0] (loopx);
\draw (aomlrx) to[in=180,out=0] (loopx);
\draw (loopx) to[in=180,out=0] (ps);
\draw (loopx) to[in=180,out=0] (bpf);

\node[draw,align=center,fill=Pastel1-E] (corx) at (10,0.5) {63 GHz\\Coherent\\Receiver};

\draw (rxamp) to[out=0,in=180] ([yshift=0.25cm]corx.west);

\node[draw,xscale=0.5,yscale=0.5,fill=Pastel1-A] (lo) at (8.5,-0.25) {\lazer};

\draw (lo.east)  to[in=180,out=0]  ([yshift=-0.25cm]corx.west);

\node[anchor=north] at (loopspool) {101.4 km};

\node[font={\Large}] at (11.25,1) {b)};




\node[draw, fill=Pastel1-A] (foe2) at (12,-0) {FOE};
\draw[-] (foe2) -- (14,-0) node[draw,fill=Pastel1-E,align=center] (mimorx) {2x2 MIMO\\65 taps};
\draw[-] (mimorx) -- (16.25,-0) node[draw,fill=Pastel1-D,align=center] (dbp2) {10 span\\$\times$10 steps};
\draw[-] (dbp2) -- (18.5,-0) node[draw,fill=Pastel1-E,align=center] (mimotx) {2x2 MIMO\\65 taps};
\draw[-] (mimotx) -- (20.5,-0) node [draw,fill=Pastel1-A] (cpe2) {CPE};
\draw[->] (cpe2) -- (21.5,-0) node[anchor=west] (snr2) {SNR};
\draw[->,Set2-D,ultra thick] (20,-1.55) .. controls (14.5,-1.55) .. (mimorx.south) node[black,midway,anchor=north] {SGD optimisation};

\draw[->,Set2-D,ultra thick] (20,-1.55) .. controls  (17,-1.55) .. (dbp2.south);
\draw[->,Set2-D,ultra thick] (20,-1.55) .. controls  (19,-1.55) .. (mimotx.south);

\draw[-,Set2-D,ultra thick] (snr2.south) .. controls (22,-1.55) .. (20,-1.55);

\end{tikzpicture}
    \caption{ a) Experimental configuration with 4$\times$30~GBd channels and 101.4~km recirculating loop. Function diagram of the receiver DSP for the L-TDDBP (b). }
    \label{fig:setup}
\end{figure*}

In this work, the deep learning of the filter weights (equivalent to a neural network's layer parameters) is implemented in Tensorflow using the RAdam optimiser\cite{Jonas2019Adaptive}. Identical complex FIR filter weights are applied to both polarisations in each step, reducing the overall number of weights to be optimized. Initialisation of the time-domain filter taps was carried out via numerical simulation of the fibre transmission link for a single channel. The forward propagation was modelled using a small  NLSE  fibre step size (100~m) at a launch power of 5~dBm (beyond optimum launch power for linear compensation). Starting from the least squares solution\cite{Eghbali2014Optimal}, a set of 10 filters was designed using a 10 span simulation. Note, these filters purely compensate fibre transmission and no transceiver impairments. Further training of the filters was carried on the experimental waveforms, before the performance was tested.

\section{Experimental Setup} 

The experimental setup is shown in Fig.~\ref{fig:setup}(a). A fibre transmission distance of 1014~km was emulated using a recirculating loop. The waveform of the 64-QAM 64-GBd channel under test (CUT) was generated offline and sent to two channels of a 33-GHz 92-GSa/s arbitrary waveform generator (AWG) and, using a dual-polarization IQ modulator (IQM), modulated the outputs of two \textless 100 kHz external cavity lasers (ECLs). Two additional 64-QAM 64-GBd aggressor channels were modulated using an additional AWG with a dual polarization IQM onto two ECLs and interleaved with the other channels to achieve uncorrelated sequences between neighbouring WDM channels. The recirculating loop with a 101.4 km span, a polarisation scrambler (PS) and three Erbium doped fibre amplifiers (EDFA) and an optical band-pass filter had the signal circulating 10 times, totalling a 1014 km transmission. At the receiver an optical band pass filter followed by an EDFA extracts the CUT for detection with a coherent receiver employing 63-GHz bandwidth 160-GS/s analogue-to-digital converters.

For this experimental demonstration, 10 steps per span were chosen for the learned TD-DBP. To take fibre loss into account, non-uniform FIR filter lengths were employed, implementing steps with equal power differences between their inputs and outputs. For the fibre nonlinearity compensation of 10 spans of 101.4 km each, the parameters were $\alpha$ of 0.16 dB/km, $\beta_2$ of -20.18 ps$^2$/km and $\gamma_\text{DBP}$ of 0.8 1/W/km. The 10 FIR filters used in the TD-DBP employed a total of 270 complex-valued tap weights at a sampling rate of 128 GSa/s. For the FD-DBP, 50 equidistant steps/span were used. This requires $2\times10\times50$ FFT operations per polarisation, while in the TD-DBP scheme the use of FFT operations is circumvented, lowering the computational complexity.

Next, for the processing of experimental data, the filter weights from simulation were used for initialisation. To prevent the dispersion filters from learning the response of the transceiver impairments, an additional 2x2 multiple input, multiple output (MIMO) filter was added before applying digital back propagation, as shown in Fig.~\ref{fig:setup}(b). Thus, the resulting structure has two linear MIMO equalisers, compensating for PMD and transmitter and receiver impairments. Note that in this way, using the automatic differentiation in Tensorflow, the filter that is applied prior the link compensation is also optimized through gradient descent. 
A root-raised cosine (RRC) filter was applied before the MIMO blocks. The carrier phase estimation was achieved by inserting pilot symbols. One in 32 symbol was a known quadrature phase shift keyed symbol (QPSK). Interpolation of the phase between the pilot symbols was performed using a Wiener filter \cite[Eq.~(32)]{Ip2007FeedForward}, following which a mean-squared-error cost is calculated.

During the training procedure, first the linear filters at both sides of the link compensation were optimised. Subsequently, all filters were updated on each optimisation step. For the FD-DBP, the $\alpha$, $\gamma$ and launched power were swept for optimisation, after which pilot-aided DSP was applied. For the experimental waveform, a single randomly generated $2^{16}$-symbol waveform was used. We split the bit sequence and corresponding received waveforms into two datasets. The first 52224 symbols (80\%) were used as training data for updating the filter weights.  The remaining 13312 symbols (20\%) were used as testing data, to obtain results reported in the figures presented.

\section{Results}
\begin{figure}[t]
    \centering
    \begin{tikzpicture}[scale=0.8,transform shape]

\definecolor{color0}{rgb}{0.12156862745098,0.466666666666667,0.705882352941177}
\definecolor{color1}{rgb}{1,0.498039215686275,0.0549019607843137}
\definecolor{color2}{rgb}{0.172549019607843,0.627450980392157,0.172549019607843}
\definecolor{color3}{rgb}{0.83921568627451,0.152941176470588,0.156862745098039}

\begin{axis}[
width = \columnwidth,
height = 0.8\columnwidth,
name=1000km,
legend cell align={left},
legend style={at={(axis cs:1,12.05)}, anchor=south, draw=white!80.0!black},
tick align=outside,
tick pos=left,
x grid style={white!69.01960784313725!black},
xlabel={Launched Power [dBm]},
xlabel style={yshift={0.15cm}},
minor tick num=1,
xmajorgrids,
grid=both,
xmin=-6, xmax=8,
xtick style={color=black},
y grid style={white!69.01960784313725!black},
ylabel={SNR [dB]},
ylabel style={yshift={-0.15cm}},
ymajorgrids,
ymin=12, ymax=16.5,
ytick style={color=black},
legend style={font=\footnotesize}, 
minor grid style={white!40!black,  dotted},
]

\addplot [semithick, color2, mark=*, mark size=3, mark options={solid}]
table {%
-6 12.8589
-5 13.3271
-4 13.8858
-3 14.5729
-2 15.0082
-1 15.5792
0 15.8163
1 15.9955
2 16.2073
3 16.2294
4 16.0649
5 15.6355
6 15.2045
7 14.4233
8 13.6074
};
\addlegendentry{learned TD-DBP 10 steps/span}


\addplot [semithick, color0, mark=square*]
table {%
-6 11.8817106189965
-5 12.6585499068657
-4 13.3031653315032
-3 13.9422837406723
-2 14.4130447865906
-1 14.9450710378497
0 15.1209815513289
1 15.4400446531188
2 15.6059976411811
3 15.8149641981244
4 15.747260094873
5 15.3726547304299
6 15.0797361134747
7 14.6194034252768
8 13.6396597422794
9 12.5328732205495
10 10.944916685203
};
\addlegendentry{FD-DBP 50-steps/span}

\addplot [semithick, color1, mark=square]
table {%
-6 11.8456350691698
-5 12.6468589964089
-4 13.2649245218054
-3 13.9152561300986
-2 14.3716319611181
-1 14.8678265339327
0 15.0472094878828
1 15.273567326118
2 15.3658726847416
3 15.3838834164162
4 15.1083737275523
5 14.5740676765246
6 13.9512590225805
7 13.2025179063993
8 12.0669554619334
9 10.8086932628224
10 9.33895861590369
};
\addlegendentry{EDC}




\coordinate (plota) at  (yticklabel cs:0.8);

\end{axis}
\node at (plota) {a)};

\end{tikzpicture}

\begin{tikzpicture}[scale=0.8,transform shape]

\definecolor{color0}{rgb}{0.12156862745098,0.466666666666667,0.705882352941177}
\definecolor{color1}{rgb}{1,0.498039215686275,0.0549019607843137}
\definecolor{color2}{rgb}{0.172549019607843,0.627450980392157,0.172549019607843}
\definecolor{color3}{rgb}{0.83921568627451,0.152941176470588,0.156862745098039}

\begin{axis}[
name=plotb,
width = \columnwidth,
height = 0.8\columnwidth,
name=1000km,
legend cell align={left},
legend style={at={(axis cs:1,12.05)}, anchor=south, draw=white!80.0!black},
tick align=outside,
tick pos=left,
x grid style={white!69.01960784313725!black},
xlabel={Launched Power [dBm]},
xlabel style={yshift={0.15cm}},
minor tick num=1,
xmajorgrids,
grid=both,
xmin=-6, xmax=8,
xtick style={color=black},
y grid style={white!69.01960784313725!black},
ylabel={SNR [dB]},
ylabel style={yshift={-0.15cm}},
ymajorgrids,
ymin=12, ymax=16.5,
ytick style={color=black},
legend style={font=\footnotesize}, 
minor grid style={white!40!black,  dotted},
]

\addplot [semithick, color2, mark=*, mark size=3, mark options={solid}]
table {%
-6 12.8589
-5 13.3271
-4 13.8858
-3 14.5729
-2 15.0082
-1 15.5792
0 15.8163
1 15.9955
2 16.2073
3 16.2294
4 16.0649
5 15.6355
6 15.2045
7 14.4233
8 13.6074
};
\addlegendentry{learned TD-DBP 10 steps/span}
\addplot [semithick, color3, mark=o, mark size=3, mark options={}]
table {%
-6 12.8588
-5 13.3158
-4 13.7074
-3 14.5626
-2 14.7783
-1 15.2199
0 15.594
1 15.6351
2 15.9427
3 15.8213
4 15.6075
5 15.0931
6 14.4617
7 13.8442
8 12.2497
};
\addlegendentry{learned ($\gamma_\text{DBP}=0$) 10 steps/span}

\addplot [semithick, Set1-D, mark=triangle, mark size=3, mark options={solid,rotate=180}]
table {%
-6 12.9236
-5 13.4129
-4 13.9732
-3 14.6498
-2 15.1635
-1 15.6918
0 15.9396
1 16.1062
2 16.3493
3 16.1566
4 16.0281
5 15.5256
6 14.5072
7 13.9863
8 12.6082
};
\addlegendentry{learned ($\gamma_\text{DBP}=0$)  1 step/link}
\draw[dashed]  (axis cs:2,15.9427) --  (axis cs:5.3,15.9427);
\draw[dashed]  (axis cs:3,16.2294) --  (axis cs:5.3,16.2294);

\draw[thick,<->] (axis cs:5,15.9427) -- (axis cs:5,16.2294) node[midway, anchor=west,fill=white,inner sep=0pt, xshift=4pt] {0.3 dB};





\coordinate (plotb) at  (yticklabel cs:0.8);

\end{axis}
\node at (plotb) {b)};

\end{tikzpicture}
    \caption{a) SNR vs. launched power for learned TD-DBP compared to conventional FD-DBP and linear EDC only methods. b) TD-DBP compaired to the same structure with $\gamma_\text{DBP}=0$ with the same number of steps and a single step for the whole link.}
    \label{fig:results}
\end{figure}
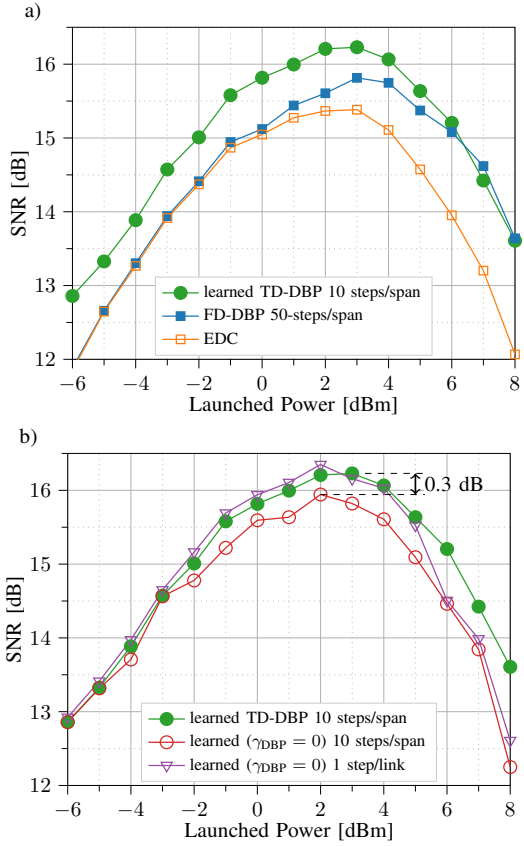

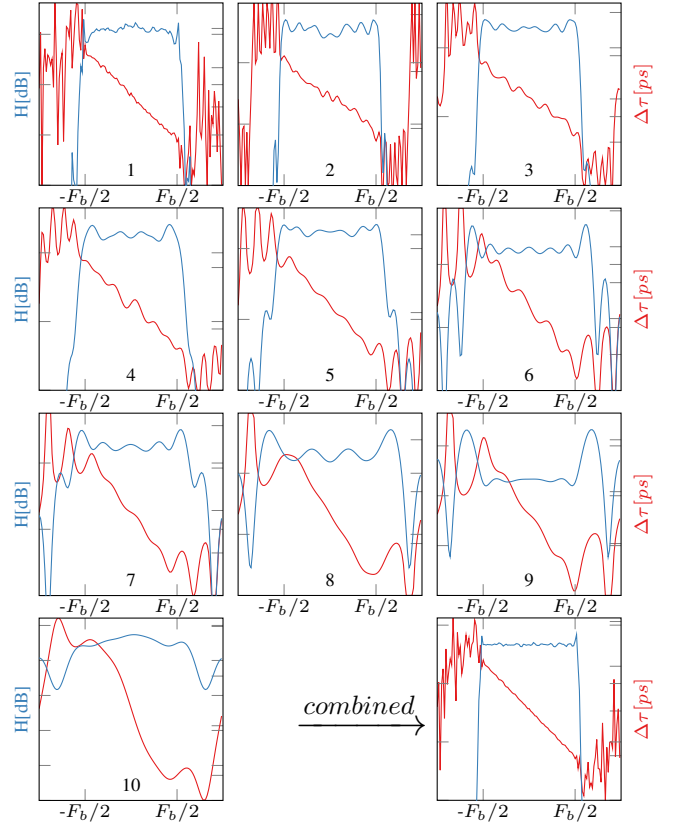
\begin{figure}[htbp]
    \centering
    \begin{tikzpicture}

\pgfplotstableread{data/exp_per_span.txt}\datatable

\begin{groupplot}[
    group style={
        group name=dplot,
        group size=3 by 4,
        ylabels at=edge right,
        horizontal sep=0.2cm,
        vertical sep=0.3cm,
    },
    height=4cm,width=4cm,
    axis y line*=right,
    yticklabel style={Set1-A,font=\scriptsize},
    yticklabels={,,},
    xtick=\empty, axis line style=transparent,
    enlarge y limits=false,
    xmin=-64,xmax=64,
    ylabel={$\Delta\tau[ps]$},ylabel style={Set1-A,yshift={0.2cm},font=\scriptsize},]
\pgfplotsinvokeforeach{11,...,20}{%
\nextgroupplot
\addplot[Set1-A] table[x index=0,y index=#1] {\datatable};
}
\nextgroupplot[group/empty plot]
\nextgroupplot
\addplot[Set1-A] table[x index=0,y index=2] {data/exp_total.txt};
\end{groupplot}
\begin{groupplot}[
    group style={
    group size=3 by 4,
        ylabels at=edge left,
        horizontal sep=0.2cm,
        vertical sep=0.3cm,
    },
    height=4cm,width=4cm,
    ymin=-10,
    xtick={-32,32},xticklabels={-${F_b}/{2}$,${F_b}/{2}$},
    xticklabel style={font=\scriptsize,yshift={0.09cm}},
    yticklabel style={Set1-B,font=\scriptsize},
    yticklabels={,,},
    xmin=-64,xmax=64,
    ylabel={H[dB]},
    ylabel style={Set1-B,yshift={-0.25cm},font=\scriptsize},
    ]

\pgfplotsinvokeforeach{1,...,10}{%
\nextgroupplot
\addplot[Set1-B] table[x index=0,y index=#1] {\datatable};
\node[anchor=south,font={\scriptsize}] at (axis cs:0,-10) {#1};
}
\nextgroupplot[group/empty plot]
\nextgroupplot
\addplot[Set1-B] table[x index=0,y index=1] {data/exp_total.txt};
\end{groupplot}
\node[align=right] at (4.3,-7) {\Large $\xrightarrow{combined}$ };

\end{tikzpicture}
    \caption{Amplitude response and group delay of the 10 individual filters used every span. Bottom right: combined response of all 10 cascaded filters.}
    \label{fig:individual_filters}
\end{figure}

The launched power was increased with 1 dB increments from -6 to +8 dBm per channel. The resulting SNR, defined as $\frac{\mathrm{E}[|X|^2]}{\mathrm{E}[|X-Y|^2]}$, where $X$ and $Y$ are the transmitted and received signal respectively. Fig.~\ref{fig:results}(a) shows a comparison of achieved SNR for TD-DBP, FD-DBP and EDC. The TD-DBP and FD-DBP are implemented using 10 and 50 steps per span respectively. Both schemes provide similar performance improvements from non-linearity compensation, with slightly higher accuracy in the high power regime for the conventional FD-DBP scheme, due to the larger number of steps used. However, the TD-DBP acheives a higher SNR in the low power regime, suggesting a better linear compensation. Fig.~\ref{fig:results}(b) compares TD-DBP with two learned linear compensation strategies. The figure shows the learned DBP performance for two cases, the proposed non-linear mitigation scheme, and the same scheme with $\gamma_\text{DBP}=0$, i.e., providing only linear compensation and a scheme where the whole chromatic dispersion is compensated in a single filter. For a low launched power into the fibre, the first two schemes show comparable performance, while a non-linearity mitigation gain of 0.3~dB is achieved at optimal launch powers. Using a single filter achieves better linear gain, but converges to the $\gamma_\text{DBP}=0$ method in the high launched power regime. 

To confirm that the algorithm is performing digital back-propagation, i.e., approximating the SSFM model, the amplitude response and group delay of the 10 individual filter used each span are plotted in Fig.~\ref{fig:individual_filters}. The expected response is an all-pass filter (H) with a linear group delay ($\Delta\tau$), compensation for chromatic dispersion. It can be seen that, while the individual filters have significant ripples, the combined filter, depicted as the last subplot of Fig.~\ref{fig:individual_filters}, has an almost perfect response within the signal bandwidth, with a flat amplitude response and a smooth group delay.

\section{Future work}

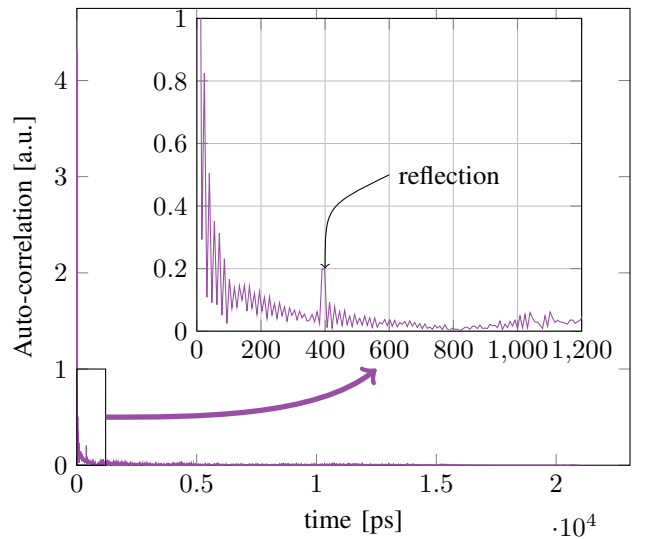
\begin{figure}
    \centering
    \begin{tikzpicture}
\begin{axis}[
width=\columnwidth,
xmin=0,ymin=0,
xlabel={time [ps]},
ylabel={Auto-correlation [a.u.]},
]
\addplot[Set1-D] table {data/filteracor.txt};

\draw (axis cs:0,0) rectangle (axis cs:1200,1);

\draw[Set1-D,->,line width=2pt] (axis cs:1200,0.5) .. controls (axis cs:5000,0.5) and  (axis cs:10000,0.5)  ..  (axis cs:12500,1); 

\coordinate (point) at (axis cs:1200,1);
\end{axis}
\begin{axis}[
width=0.75\columnwidth, at=(point), xshift=1.2cm,yshift=0.5cm,
xmin=0,xmax=1200,
ymin=0,ymax=1,
grid=major,
]
\addplot[Set1-D] table {data/filteracor.txt};

\draw[->] (axis cs:600,0.5) node[anchor=west] {reflection} .. controls (axis cs:400,0.4) .. (axis cs:400,0.2);
\end{axis}
\end{tikzpicture}
    \caption{The auto-correlation of the learned single filter used to compensate chromatic dispersion.}
    \label{fig:filteracor}
\end{figure}

We have also trained a single convolutional layer for the whole link to apply the chromatic dispersion compensation and shown the results in Fig.~\ref{fig:results}. This method outperforms all other methods up until optimal launched power. However for high launched powers the performance converge to the linear compensation results. When looking at the auto-correlation of the single learned filter in Fig.~\ref{fig:filteracor}, we can see bumps where the filter has compensated a reflection. This suggests that not all the linear effects are compensated for and combining this result with nonlinear compensation will increase the performance even further.

We suspect the difference in performance between the single layer and the deep network to partially be attributable to gradient propagation through the many layers, in this work over 100. A method to combat this is proposed in \cite{He2015Deep}, where residual links are bypassing the layer. The result are equivalent due to the universal function approximates used. When trying to learn $x_{k+1}=\mathcal{U}\{x_k\}$ will be equivalent to learning $x_{k+1}=x_k+\mathcal{V}\{x_k\}$ if $\mathcal{V}\{x\} \triangleq \mathcal{U}\{x\}-x$. These layers have the gradient
\begin{align}
    \frac{\partial x_{k+1}}{\partial x_k} &= \mathcal{U}^\prime\{x_k\} \\
    \frac{\partial x_{k+1}}{\partial x_k} &= 1 + \mathcal{V}^\prime \{x_k\}
\end{align}
but if two layers are applied sequentially, the gradient becomes
\begin{align}
    x_{k+2} &= \mathcal{U}\{\mathcal{U}\{x_k\}\}\\
    \frac{\partial x_{k+2}}{\partial x_k} &= \mathcal{U}^\prime\{x_{k+1}\}\mathcal{U}^\prime\{x_k\} \\
    \frac{\partial x_{k+2}}{\partial x_k} &= 1 + \mathcal{V}^\prime \{x_{k+1}\}\left(1+\mathcal{V}^\prime\{x_k\}\right)
\end{align}
We can see the direct function will have a multiplicative term for every layer, but the residual link will propagate the constant through all the layers and therefore a more direct link with the error function.

For the DPB we can see this calculation the chromatic dispersion and nonlinear phase shift as perturbation to our signal. This is a different approach then the perturbation DBP \cite{Liang2014Multi-stage,Yan2011Low}, although it shares some thoughts.

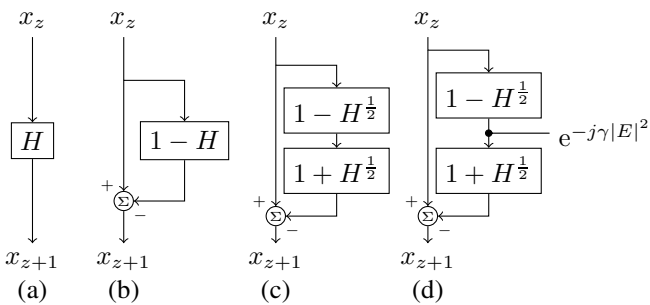
\begin{figure}[h]
    \centering
    \begin{tikzpicture}[scale=0.8]
    \node (in1) at (0,0) {$x_z$};
    \node[draw] (h1) at (0,-2) {$H$};
    \node (out1) at (0,-4) {$x_{z+1}$};
    \draw[->] (in1) -- (h1);
    \draw[->] (h1) -- (out1);

    \node (in2) at (1.5,0) {$x_z$};
    \coordinate (split2) at (1.5,-1);
    \node[draw] (h2) at (2.5,-2) {$1-H$};
    \node[inner sep=0.5pt,draw,circle]  (plus2) at (1.5,-3) {\tiny $\Sigma$};
    \node (out2) at (1.5,-4) {$x_{z+1}$};
    \draw[->] (in2) -- (split2) --+(1,0) -- (h2);
    \draw[->] (h2) -- +(0,-1) -- (plus2) node[anchor=north west] {\tiny$-$};
    \draw[->] (split2) -- (plus2) node[anchor=south east] {\tiny$+$};
    \draw[->] (plus2) -- (out2);

    \node (in3) at (4,0) {$x_z$};
    \coordinate (split3) at (4,-0.75);
    \node[draw] (h3m) at (5,-1.5) {$1-H^{\frac12}$};
    \node[draw] (h3p) at (5,-2.5) {$1+H^{\frac12}$};
    \node[inner sep=0.5pt,draw,circle] (plus3) at (4,-3.25) {\tiny $\Sigma$};
    \node (out3) at (4,-4) {$x_{z+1}$};
    \draw[->] (in3) -- (split3) -- +(1,0) -- (h3m);
    \draw[->] (h3m) -- (h3p);
    \draw[->] (h3p) -- +(0,-0.75) -- (plus3) node[anchor=north west] {\tiny$-$};
    \draw[->] (split3) -- (plus3) node[anchor=south east] {\tiny$+$};
    \draw[->] (plus3) -- (out3);

    \node (in4) at (6.5,0) {$x_z$};
    \coordinate (split4) at (6.5,-0.5);
    \node[draw] (h4m) at (7.5,-1.25) {$1-H^{\frac12}$};
    \node[draw] (h4p) at (7.5,-2.5) {$1+H^{\frac12}$};
    \node[inner sep=0.5pt,draw,circle] (plus4) at (6.5,-3.25) {\tiny $\Sigma$};
    \node (out4) at (6.5,-4) {$x_{z+1}$};
    \draw[->] (in4) -- (split4) -- +(1,0) -- (h4m);
    \draw[->] (h4m) -- (h4p);
    \draw[->] (h4p) -- +(0,-0.75) -- (plus4) node[anchor=north west] {\tiny$-$};
    \draw[->] (split4) -- (plus4) node[anchor=south east] {\tiny$+$};
    \draw[->] (plus4) -- (out4);
    \draw[fill=black] (7.5,-1.875) circle(1.5pt) -- +(1,0) node[anchor=west] {$\mathrm{e}^{-j\gamma|E|^2}$};
    
    \node at (0,-4.5) {(a)};
    \node at (1.5,-4.5) {(b)};
    \node at (4,-4.5) {(c)};
    \node at (6.5,-4.5) {(d)};

\end{tikzpicture}
    \caption{The proposed digital back-propagation block. (a) a single dispersion block (b) add the residual link (c) split the block into to half steps (d) add the nonlinear phase shift.}
    \label{fig:resnet}
\end{figure}

In Fig.~\ref{fig:resnet}, the proposed block with the residual link is shown (d). The block is designed from starting with a dispersion block in the frequency domain $x_{k+1} = H(x) = x\mathrm{e}^{-jK(\omega T)^2}$, which can either be applied directly (a) or with a residual link (b). However, in (c) we have split the dispersion step such that we have 
\begin{align}
    x_{k+1} &= x_k - x_k\left(1-\mathrm{e}^\frac{-jK(\omega T)^2}{2}\right)\left(1+\mathrm{e}^\frac{-jK(\omega T)^2}{2}\right)\nonumber \\
     &= x_k - x_k\Bigg(1 -\left(\mathrm{e}^\frac{-jK(\omega T)^2}{2}\right)^2\Bigg) \nonumber\\
     &= x_k\mathrm{e}^{-jK(\omega T)^2}
\end{align}

Now we have split the step, we can apply the nonlinear phase shift in the middle of the block. Effectively recreating a split step method with a residual link.

\section{Conclusion}
We experimentally demonstrated learned time-domain digital back-propagation for the first time. The learned algorithm was verified to approximate the NLSE model, showing flat amplitude response and smooth group delay for the cascaded filters. Overall, an SNR improvement due to non-linearity mitigation of 0.3 dB was achieved, comparable with conventional mitigation algorithms.

{
\small
\textit{Financial support by EPSRC TRANSNET and EU COIN. E. Sillekens funded by EPSRC grant EP/M507970/1.}\vspace{-0.3cm}
}

\bibliographystyle{IEEEtran}
\bibliography{sample}

\end{document}